\pdfoutput=1
\documentclass[12pt,a4paper]{article}

\usepackage{ifthen} 
\newboolean{pdflatex}
\setboolean{pdflatex}{true} 

\newboolean{articletitles}
\setboolean{articletitles}{true} 

\newboolean{uprightparticles}
\setboolean{uprightparticles}{false} 

\def\paperauthors{LHCb collaboration} 
\def\paperasciititle{First evidence of the decay B->piee} 
\def\papertitle{First evidence of the decay $B^+\to\pi^+ e^+ e^-$} 
\def\paperkeywords{{High Energy Physics}, {LHCb}} 
\def\papercopyright{\the\year\ CERN for the benefit of the LHCb collaboration} 
\def\paperlicence{CC BY 4.0 licence}
\def\paperlicenceurl{https://creativecommons.org/licenses/by/4.0/}

\newif\ifEnableSectionTOCLinks
\EnableSectionTOCLinksfalse 

\usepackage[top=1in, bottom=1.25in, left=1in, right=1in]{geometry}

\usepackage{microtype}
\usepackage{lineno}  
\usepackage{xspace} 
\usepackage{caption} 

\usepackage{graphicx}  
\usepackage{color}
\usepackage{colortbl}
\graphicspath{{./figs/}} 

\usepackage{amsmath} 
\usepackage{amssymb}
\usepackage{amsfonts}
\usepackage{upgreek} 

\newcommand*\patchAmsMathEnvironmentForLineno[1]{%
\expandafter\let\csname old#1\expandafter\endcsname\csname #1\endcsname
\expandafter\let\csname oldend#1\expandafter\endcsname\csname
end#1\endcsname
 \renewenvironment{#1}%
   {\linenomath\csname old#1\endcsname}%
   {\csname oldend#1\endcsname\endlinenomath}%
}
\newcommand*\patchBothAmsMathEnvironmentsForLineno[1]{%
  \patchAmsMathEnvironmentForLineno{#1}%
  \patchAmsMathEnvironmentForLineno{#1*}%
}
\AtBeginDocument{%
\patchBothAmsMathEnvironmentsForLineno{equation}%
\patchBothAmsMathEnvironmentsForLineno{align}%
\patchBothAmsMathEnvironmentsForLineno{flalign}%
\patchBothAmsMathEnvironmentsForLineno{alignat}%
\patchBothAmsMathEnvironmentsForLineno{gather}%
\patchBothAmsMathEnvironmentsForLineno{multline}%
\patchBothAmsMathEnvironmentsForLineno{eqnarray}%
}

\usepackage[pdftex,
            pdfauthor={\paperauthors},
            pdftitle={\paperasciititle},
            pdfkeywords={\paperkeywords}]{hyperref}
\usepackage{hyperxmp}
\hypersetup{
    pdfcopyright={Copyright (C) \papercopyright},
    pdflicenseurl={\paperlicenceurl}
}

\usepackage[colorinlistoftodos,textsize=scriptsize]{todonotes}

\usepackage[bottom,flushmargin,hang,multiple]{footmisc}

\usepackage[all]{hypcap} 

\usepackage{xspace}
\usepackage{upgreek}

\def\lhcb   {\mbox{LHCb}\xspace}

\def\belle  {\mbox{Belle}\xspace}
\def\belletwo {\mbox{Belle~II}\xspace}

\def\MagUp {\mbox{\em Mag\kern -0.05em Up}\xspace}

\ifthenelse{\boolean{uprightparticles}}%
{

 \def\Pmu         {\ensuremath{\upmu}\xspace}
 \def\Pnu         {\ensuremath{\upnu}\xspace}
 
 \def\Ppi         {\ensuremath{\uppi}\xspace}

 \def\Ppsi        {\ensuremath{\uppsi}\xspace}

 \def\PDelta      {\ensuremath{\Delta}\xspace}
 \def\PXi         {\ensuremath{\Xi}\xspace}
 \def\PLambda     {\ensuremath{\Lambda}\xspace}
 \def\PSigma      {\ensuremath{\Sigma}\xspace}
 \def\POmega      {\ensuremath{\Omega}\xspace}
 \def\PUpsilon    {\ensuremath{\Upsilon}\xspace}
 \let\oldPi\Pi
 \def\PPi         {\ensuremath{\oldPi}\xspace}

 \def\PB      {\ensuremath{\mathrm{B}}\xspace}
 \def\PD      {\ensuremath{\mathrm{D}}\xspace}
 \def\PJ      {\ensuremath{\mathrm{J}}\xspace}
 \def\PK      {\ensuremath{\mathrm{K}}\xspace}
 \def\Pb      {\ensuremath{\mathrm{b}}\xspace}
 \def\Pc      {\ensuremath{\mathrm{c}}\xspace}
 \def\Pd      {\ensuremath{\mathrm{d}}\xspace}
 \def\Pe      {\ensuremath{\mathrm{e}}\xspace}
 \def\Ps      {\ensuremath{\mathrm{s}}\xspace}
 \def\Pt      {\ensuremath{\mathrm{t}}\xspace}

 \def\thebaroffset{0.0em}
}
{

 \def\Pmu         {\ensuremath{\mu}\xspace}
 \def\Pnu         {\ensuremath{\nu}\xspace}
 
 \def\Ppi         {\ensuremath{\pi}\xspace}

 \def\Ppsi        {\ensuremath{\psi}\xspace}
 
 \mathchardef\PDelta="7101
 \mathchardef\PXi="7104
 \mathchardef\PLambda="7103
 \mathchardef\PSigma="7106
 \mathchardef\POmega="710A
 \mathchardef\PUpsilon="7107
 \mathchardef\PPi="7105
 \def\PB      {\ensuremath{B}\xspace}
 \def\PD      {\ensuremath{D}\xspace}
 \def\PJ      {\ensuremath{J}\xspace}
 \def\PK      {\ensuremath{K}\xspace}
 \def\Pb      {\ensuremath{b}\xspace}
 \def\Pc      {\ensuremath{c}\xspace}
 \def\Pd      {\ensuremath{d}\xspace}
 \def\Pe      {\ensuremath{e}\xspace}
 \def\Ps      {\ensuremath{s}\xspace}
 \def\Pt      {\ensuremath{t}\xspace}

 \def\thebaroffset{0.18em}
}
\newcommand{\offsetoverline}[2][\thebaroffset]{\kern #1\overline{\kern -#1 #2}}%

\makeatletter
\ifcase \@ptsize \relax
  \newcommand{\miniscule}{\@setfontsize\miniscule{4}{5}}
\or
  \newcommand{\miniscule}{\@setfontsize\miniscule{5}{6}}
\or
  \newcommand{\miniscule}{\@setfontsize\miniscule{5}{6}}
\fi
\makeatother

\DeclareRobustCommand{\optbar}[1]{\shortstack{{\miniscule (\rule[.5ex]{1.25em}{.18mm})}
  \\ [-.7ex] $#1$}}

\def\en         {{\ensuremath{\Pe^-}}\xspace}   
\def\ep         {{\ensuremath{\Pe^+}}\xspace}

\def\epem       {{\ensuremath{\Pe^+\Pe^-}}\xspace}

\def\mup        {{\ensuremath{\Pmu^+}}\xspace}
\def\mun        {{\ensuremath{\Pmu^-}}\xspace} 

\def\ellm       {{\ensuremath{\ell^-}}\xspace}
\def\ellp       {{\ensuremath{\ell^+}}\xspace}

\def\neu        {{\ensuremath{\Pnu}}\xspace}
\def\neub       {{\ensuremath{\overline{\Pnu}}}\xspace}
\def\neue       {{\ensuremath{\neu_e}}\xspace}
\def\neueb      {{\ensuremath{\neub_e}}\xspace}

\def\dquark    {{\ensuremath{\Pd}}\xspace}

\def\squark    {{\ensuremath{\Ps}}\xspace}

\def\cquark    {{\ensuremath{\Pc}}\xspace}

\def\bquark    {{\ensuremath{\Pb}}\xspace}

\def\tquark    {{\ensuremath{\Pt}}\xspace}

\def\pion   {{\ensuremath{\Ppi}}\xspace}

\def\pip    {{\ensuremath{\pion^+}}\xspace}

\def\kaon    {{\ensuremath{\PK}}\xspace}

\def\KorKbar {\kern \thebaroffset\optbar{\kern -\thebaroffset \PK}{}\xspace}

\def\Kp      {{\ensuremath{\kaon^+}}\xspace}

\def\Dbar    {{\ensuremath{\offsetoverline{\PD}}}\xspace}
\def\D       {{\ensuremath{\PD}}\xspace}

\def\DorDbar {\kern \thebaroffset\optbar{\kern -\thebaroffset \PD}\xspace}
\def\Dz      {{\ensuremath{\D^0}}\xspace}
\def\Dzb     {{\ensuremath{\Dbar{}^0}}\xspace}
\def\Dp      {{\ensuremath{\D^+}}\xspace}
\def\Dm      {{\ensuremath{\D^-}}\xspace}

\def\DpDm    {\ensuremath{\Dp {\kern -0.16em \Dm}}\xspace}

\def\B       {{\ensuremath{\PB}}\xspace}

\def\BorBbar {\kern \thebaroffset\optbar{\kern -\thebaroffset \PB}\xspace}

\def\Bd      {{\ensuremath{\B^0}}\xspace}

\def\BdorBdbar {\kern \thebaroffset\optbar{\kern -\thebaroffset \Bd}\xspace}
\def\Bu      {{\ensuremath{\B^+}}\xspace}

\def\Bp      {{\ensuremath{\Bu}}\xspace}

\def\Bs      {{\ensuremath{\B^0_\squark}}\xspace}

\def\BsorBsbar {\kern \thebaroffset\optbar{\kern -\thebaroffset \Bs}\xspace}

\def\jpsi     {{\ensuremath{{\PJ\mskip -3mu/\mskip -2mu\Ppsi}}}\xspace}
\def\psitwos  {{\ensuremath{\Ppsi{(2S)}}}\xspace}

\def\Y#1S{\ensuremath{\PUpsilon{(#1S)}}\xspace}

\def\LorLbar     {\kern \thebaroffset\optbar{\kern -\thebaroffset \PLambda}\xspace}

\def\BF         {{\ensuremath{\mathcal{B}}}\xspace}
\def\BR         {\BF}

\newcommand{\decay}[2]{\mbox{\ensuremath{#1\!\to #2}}\xspace}

\def\to                 {\ensuremath{\rightarrow}\xspace}

\def\qsq       {{\ensuremath{q^2}}\xspace}

\def\CP                {{\ensuremath{C\!P}}\xspace}

\def\Vtd  {{\ensuremath{V_{\tquark\dquark}^{\phantom{\ast}}}}\xspace}

\def\Vts  {{\ensuremath{V_{\tquark\squark}^{\phantom{\ast}}}}\xspace}

\def\AT#1     {\ensuremath{A_{\mathrm{T}}^{#1}}\xspace}           

\def\C#1      {\ensuremath{\mathcal{C}_{#1}}\xspace}                       
\def\Cp#1     {\ensuremath{\mathcal{C}_{#1}^{'}}\xspace}                    
\def\Ceff#1   {\ensuremath{\mathcal{C}_{#1}^{\mathrm{(eff)}}}\xspace}        
\def\Cpeff#1  {\ensuremath{\mathcal{C}_{#1}^{'\mathrm{(eff)}}}\xspace}       
\def\Ope#1    {\ensuremath{\mathcal{O}_{#1}}\xspace}                       
\def\Opep#1   {\ensuremath{\mathcal{O}_{#1}^{'}}\xspace}                    

\newcommand{\aunit}[1]{\ensuremath{\text{\,#1}}}

\newcommand{\tev}{\aunit{Te\kern -0.1em V}\xspace}
\newcommand{\gev}{\aunit{Ge\kern -0.1em V}\xspace}
\newcommand{\mev}{\aunit{Me\kern -0.1em V}\xspace}
\newcommand{\kev}{\aunit{ke\kern -0.1em V}\xspace}
\newcommand{\ev}{\aunit{e\kern -0.1em V}\xspace}

\newcommand{\mevc}{\ensuremath{\aunit{Me\kern -0.1em V\!/}c}\xspace}
\newcommand{\gevc}{\ensuremath{\aunit{Ge\kern -0.1em V\!/}c}\xspace}
\newcommand{\mevcc}{\ensuremath{\aunit{Me\kern -0.1em V\!/}c^2}\xspace}
\newcommand{\gevcc}{\ensuremath{\aunit{Ge\kern -0.1em V\!/}c^2}\xspace}
\newcommand{\gevgevcccc}{\ensuremath{\gev^2\!/c^4}\xspace} 

\def\fb   {\ensuremath{\aunit{fb}}\xspace}
\def\invfb   {\ensuremath{\fb^{-1}}\xspace}

\def\gsim{{~\raise.15em\hbox{$>$}\kern-.85em
          \lower.35em\hbox{$\sim$}~}\xspace}
\def\lsim{{~\raise.15em\hbox{$<$}\kern-.85em
          \lower.35em\hbox{$\sim$}~}\xspace}

\def\pt         {\ensuremath{p_{\mathrm{T}}}\xspace}

\def\evtgen     {\mbox{\textsc{EvtGen}}\xspace}

\def\geant      {\mbox{\textsc{Geant4}}\xspace}

\def\photos     {\mbox{\textsc{Photos}}\xspace}

\def\pythia     {\mbox{\textsc{Pythia}}\xspace}

\def\tell1  {TELL1\xspace}
\def\ukl1   {UKL1\xspace}

\newcommand{\ie}{\mbox{\itshape i.e.}\xspace}

\newcommand{\lhcborcid}[1]{\href{https://orcid.org/#1}{\hspace*{0.1em}\raisebox{-0.45ex}{\includegraphics[width=1em]{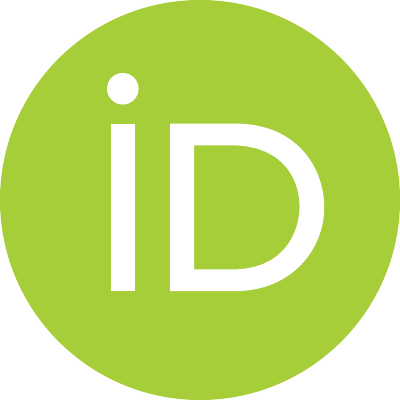}}}}

\def\BuTopill {\decay{\Bu}{\pip \ellp \ellm}}
\def\BuTopiee {\decay{\Bu}{\pip \ep \en}}
\def\BuTopimm {\decay{\Bu}{\pip \mup \mun}}

\def\BuTopiJpsi {\decay{\Bu}{\pip \jpsi}}
\def\BuTopiJpsiee {\decay{\Bu}{\pip \jpsi \left(\ep \en \right)}}

\def\BuToKee {\decay{\Bu}{\Kp \ep \en}}

\def\BuTopsitwosee {\decay{\Bu}{\pip\psitwos(\epem)}}

\def\AT#1     {\ensuremath{A_{\mathrm{T}}^{#1}}\xspace}           

\def\BuToKJpsi {\decay{\Bu}{\Kp \jpsi }}
\def\BuToKJpsiee {\decay{\Bu}{\Kp \jpsi \left(\epem \right)}}
\def\BuToKJpsimm {\decay{\Bu}{\Kp \jpsi \left(\mup\mun \right)}}

\def\C#1      {\ensuremath{\mathcal{C}_{#1}}\xspace}                       
\def\Cp#1     {\ensuremath{\mathcal{C}_{#1}^{'}}\xspace}                    
\def\Ceff#1   {\ensuremath{\mathcal{C}_{#1}^{\mathrm{(eff)}}}\xspace}        
\def\Cpeff#1  {\ensuremath{\mathcal{C}_{#1}^{'\mathrm{(eff)}}}\xspace}       
\def\Ope#1    {\ensuremath{\mathcal{O}_{#1}}\xspace}                       
\def\Opep#1   {\ensuremath{\mathcal{O}_{#1}^{'}}\xspace}                    

\def\qsq            {\ensuremath{q^2}\xspace}

\newcommand{\qsqtrack}{\ensuremath{q^2_{\mathrm{track}}}\xspace}

\newcommand{\figref}[1]{Fig.~\ref{#1}}

\newcommand{\tabref}[1]{Table~\ref{#1}}

\newcommand{\mpiee}{\ensuremath{m(\pi^+e^+e^-)}\xspace}
\def\BuTorhopzee {\decay{B^{+,0}}{\rho(770)^{+,0} e^+ e^-}}
\def\BdToKstpzee {\decay{B^{+,0}}{K^{*} (892)^{+,0} e^+ e^-}}

\def\BuToXpsitwosee {\decay{B}{X \psitwos\left(\ep \en \right)}}
\def\BuToXJpsiee {\decay{B}{X \jpsi\left(\ep \en \right)}}

\hypersetup{
  colorlinks   = true, 
  urlcolor     = blue, 
  linkcolor    = blue, 
  citecolor    = red   
}

\ifEnableSectionTOCLinks
    \usepackage[explicit]{titlesec} 
    
    \let\oldcontentsline\contentsline
    \renewcommand

    \titleformat{\section}{\normalfont\Large\bf}{\hyperlink{tocsection.\thesection}{{\thesection} \parbox[t]{\dimexpr\textwidth-1pc}{#1}}}{1pc}{}

    \titleformat{\subsection}{\normalfont\bf}{\hyperlink{tocsubsection.\thesubsection}{{\thesubsection} \parbox[t]{\dimexpr\textwidth-1pc}{#1}}}{1pc}{}

    \titleformat{name=\section,numberless}[display]{}{}{0pt}{\normalfont\Huge\bfseries #1}
\fi

\usepackage{cite} 
\usepackage{mciteplus}

\usepackage{longtable} 

\begin{document}

\renewcommand{\thefootnote}{\fnsymbol{footnote}}
\setcounter{footnote}{1}


\begin{titlepage}
\pagenumbering{roman}

\vspace*{-1.5cm}
\centerline{\large EUROPEAN ORGANIZATION FOR NUCLEAR RESEARCH (CERN)}
\vspace*{1.5cm}
\noindent
\begin{tabular*}{\linewidth}{lc@{\extracolsep{\fill}}r@{\extracolsep{0pt}}}
\ifthenelse{\boolean{pdflatex}}
{\vspace*{-1.5cm}\mbox{\!\!\!\includegraphics[width=.14\textwidth]{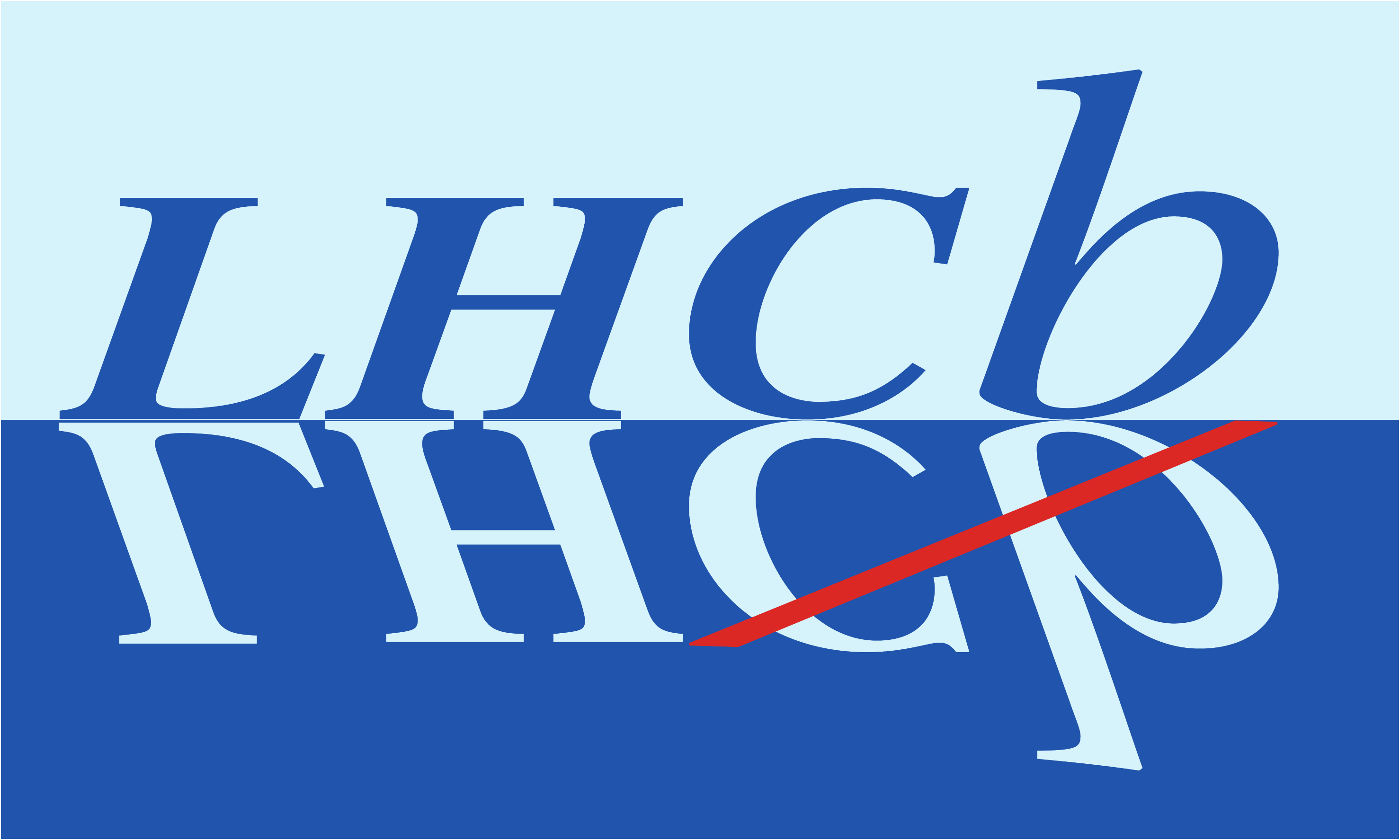}} & &}%
{\vspace*{-1.2cm}\mbox{\!\!\!\includegraphics[width=.12\textwidth]{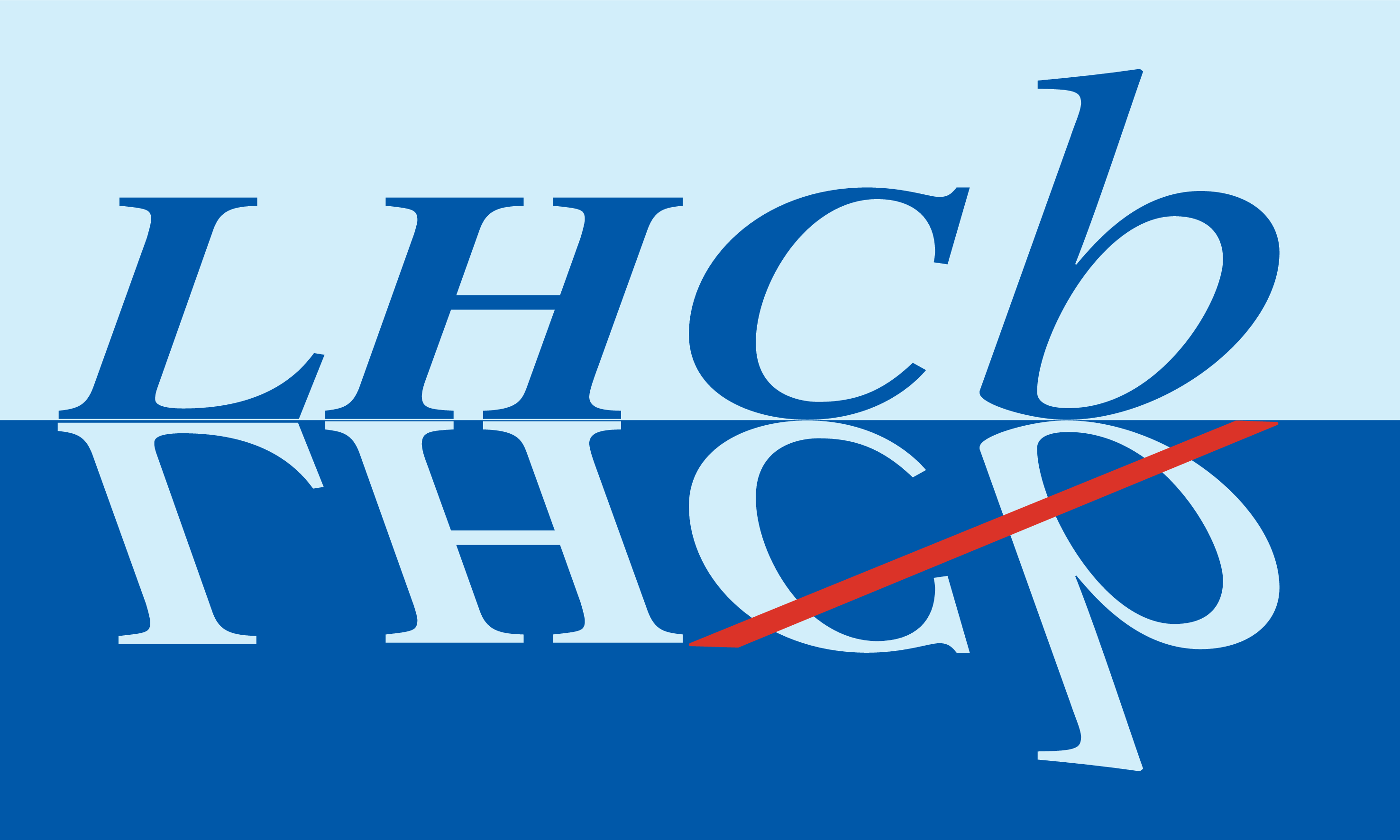}} & &}%
\\
 & & CERN-EP-2026-055 \\  
 & & LHCb-PAPER-2025-050 \\  
 & & October 26, 2026 \\ 
 & & \\
\end{tabular*}

\vspace*{4.0cm}

{\normalfont\bfseries\boldmath\huge
\begin{center}
  \papertitle 
\end{center}
}

\vspace*{2.0cm}

\begin{center}
\paperauthors\footnote{Authors are listed at the end of this Letter.}
\end{center}

\vspace{\fill}

\begin{abstract}
  \noindent
  The first evidence for the decay $B^+\to\pi^+ e^+ e^-$ is reported using proton-proton collision data recorded by the LHCb experiment at centre-of-mass energies of 7, 8 and 13 TeV, corresponding to an integrated luminosity of 9 fb$^{-1}$. A signal excess with a significance of 3.2$\sigma$ is observed and the branching fraction is measured to be $\mathcal{B}(B^+\to\pi^+ e^+ e^-) = (2.4\,{}^{+0.9}_{-0.8} \,{}^{+0.4}_{-0.2}) \times 10^{-8}$, where the first set of uncertainties is statistical and the second is systematic. The result is consistent with the Standard Model expectation.
\end{abstract}

\vspace*{2.0cm}

\begin{center}
  Published in Phys.~Rev.~Lett. 137, (2026) 071804.

\end{center}

\vspace{\fill}

{\footnotesize 
\centerline{\copyright~\papercopyright. \href{\paperlicenceurl}{\paperlicence}.}}
\vspace*{2mm}

\end{titlepage}


\newpage
\setcounter{page}{2}
\mbox{~}
%
%
%
%


\renewcommand{\thefootnote}{\arabic{footnote}}
\setcounter{footnote}{0}


\cleardoublepage


\pagestyle{plain} 
\setcounter{page}{1}
\pagenumbering{arabic}



Flavour-changing neutral-current (FCNC) decays of $b$ hadrons are suppressed in the Standard Model (SM) and therefore constitute a good laboratory to search for the effects of new physics. Several tensions with respect to the SM predictions have been observed in the measurements of branching fractions~\cite{LHCb-PAPER-2025-055, LHCb-PAPER-2021-014, LHCb-PAPER-2014-006} and angular observables~\cite{LHCb-PAPER-2025-041, CMS:2024atz, LHCb-PAPER-2024-011, LHCb-PAPER-2023-033, LHCb-PAPER-2023-032, LHCb-PAPER-2021-022, LHCb-PAPER-2020-041, ATLAS:2018gqc, Belle:2016fev} of decays mediated by the $\decay{b}{s \ell \ell}$ transition, whereas the most recent tests of lepton flavour universality in these processes are consistent with the SM~\cite{LHCb-PAPER-2022-046, LHCb-PAPER-2024-056,LHCb-PAPER-2024-032}. 
The decay \BuTopill is mediated by the $\decay{b}{d\ell\ell}$ quark-level transition, which is further suppressed by the Cabibbo–Kobayashi–Maskawa (CKM) matrix element ratio $|\Vtd|^2/|\Vts|^2$ in the SM and is also sensitive to new physics effects. Furthermore, the additional CKM suppression might be absent in new physics scenarios where the flavour structure of the SM is not preserved. These could lead to modifications of the branching fraction, \CP asymmetries and lepton flavour universality observables~\cite{Khodjamirian_pill, piee_SM, Bordone:2021olx} in these processes.   

The \BuTopimm decay was first observed by LHCb~\cite{LHCB-PAPER-2012-020}, and its total branching fraction was measured to be
$\BR(\BuTopimm) = (1.83 \pm 0.25 ) \times 10^{-8}$~\cite{LHCb-PAPER-2015-035},
in agreement with the SM prediction of $\BR(\BuTopill) = (2.04 \pm 0.21 ) \times 10^{-8}$ ~\cite{piee_SM}. Evidence for other $\decay{b}{d \mu^+ \mu^-}$ decays has also been reported~\cite{LHCB-PAPER-2018-004, LHCB-PAPER-2016-049, LHCB-PAPER-2014-063}, while no $\decay{b}{d e^+ e^-}$ quark-level transition has been observed so far.
The most sensitive searches come from the \belle and \belletwo experiments, which reported an upper limit of $\BR<5.4 \times 10^{-8}$ at 90\% confidence level for the \BuTopiee decay~\cite{Belle_piee}.

In this Letter, a search for the \BuTopiee decay is presented using data from $pp$ collisions at centre-of-mass energies of $\sqrt{s}$ = 7, 8 and 13 \tev, corresponding to an integrated luminosity of 9\invfb, recorded by the LHCb experiment during Run~1 (2011\textendash2012) and Run~2 (2015\textendash2018) of the LHC. The \BuTopiee branching fraction is measured, using the \BuToKJpsiee decay as normalisation channel. The inclusion of charge-conjugate processes is implied throughout this Letter.

The LHCb detector~\cite{LHCb-DP-2008-001,LHCb-DP-2014-002} is a single-arm forward spectrometer covering the pseudorapidity range $2<\eta <5$,  designed for the study of particles containing \bquark or \cquark quarks. The online event selection is performed by a trigger, consisting of a hardware stage based on information from the calorimeter and muon systems, followed by a software stage which applies a full event reconstruction. In this measurement, the hardware trigger must be satisfied by a cluster in the electromagnetic calorimeter (ECAL) with high energy transverse to the beamline that is associated with at least one of the electrons forming the signal candidate, or by particles in the event not used to reconstruct the signal candidate. The first stage of the software trigger requires at least one charged particle with a large transverse momentum (\pt), which is inconsistent with originating from any primary $pp$ interaction vertex (PV). The second stage requires a two- or three-track secondary vertex with a significant displacement from any PV, where a multivariate algorithm~\cite{BBDT,LHCb-PROC-2015-018} is used to identify secondary vertices consistent with \bquark-hadron decays.

Simulated $pp$ collisions are generated using \pythia~\cite{Sjostrand:2007gs} with a specific \lhcb configuration~\cite{LHCb-PROC-2010-056}. Decays of unstable particles are described by \evtgen~\cite{Lange:2001uf}, in which final-state radiation is generated using \photos~\cite{davidson2015photos}. The interaction of the generated particles with the detector, and its response, are implemented using the \geant toolkit~\cite{Allison:2006ve, *Agostinelli:2002hh} as described in Ref.~\cite{LHCb-PROC-2011-006}. Simulated signal decays are generated with a dielectron-mass-squared distribution following the SM prediction in Ref.~\cite{Ball:2004ye}. Furthermore, the generated \pt and pseudorapidity distributions of the $B^+$ meson, as well as the multiplicity of the $pp$ collisions, are corrected for discrepancies between data and simulation using samples of $\BuToKJpsimm$ decays.


Candidate \BuTopiee and \BuToKJpsiee decays are reconstructed from a particle identified as a charged hadron (pion or kaon, respectively), together with a pair of well-reconstructed oppositely charged particles identified as electrons. Kaons, pions and electrons are identified using multivariate classifiers that employ particle-identification (PID) information from the two ring-imaging Cherenkov (RICH) detectors, the calorimeters and the muon chambers. Electron, pion and kaon candidates are also required to have sizeable \pt and to be inconsistent with originating from a PV. The particles must form a good-quality common vertex that is significantly displaced from all PVs. The \Bp momentum vector is required to be aligned with the vector connecting one of the PVs in the event and the \Bp decay vertex. 

Electrons lose a significant fraction of their energy through bremsstrahlung in the detector. To account for this effect, the electron momentum is corrected to include the energy of any reconstructed clusters in the ECAL that match the position of the extrapolated electron track before the LHCb dipole magnet and are not associated with any other track in the event. The invariant-mass resolution and the PID selection efficiency differ for signal candidates where photon clusters have been added to the candidate by the bremsstrahlung recovery algorithm (HasBrem) and those for which no clusters were found (NoBrem). In particular, background sources where one or more hadrons are misidentified as electrons are negligible in the HasBrem category, but significant in the NoBrem category. Therefore, these two categories are considered separately.

Signal candidates are considered in two regions of dielectron mass squared, \qsq: a low-\qsq region, $q^2\in[0.045,6]\gevgevcccc$, which is above the region contaminated by photon-to-electron conversions but below the charmonium resonances, and a high-\qsq region, $q^2\in[15,25]\gevgevcccc$, above the $\psi(2S)$ resonance. Background \BuTopsitwosee decays where spurious photon clusters have been added are removed from the high-\qsq region by requiring that the dielectron mass squared calculated before any bremsstrahlung photon is added, \qsqtrack, is more than $14\gevgevcccc$ for candidates with $q^2\in[15,18]\gevgevcccc$. The \BuToKJpsiee normalisation mode is selected in the range \mbox{$q^2\in[6,11]\gevgevcccc$}.

As a consequence of the low expected signal branching fraction, requirements are imposed to reject multiple kinds of background processes.
Most relevant are fully hadronic \mbox{$\decay{B^+}{h^+h^-h^+}$} decays, where two hadrons are misidentified as electrons, since they have branching fractions several orders of magnitude larger than the signal and peak sharply in the same region of invariant mass. These background contributions, along with semileptonic $\decay{B^+}{h^+ h^- \ep \neue}$ decays, are suppressed to manageable levels with stringent electron PID requirements.

Backgrounds where a kaon is misidentified as a pion, from both fully and partially reconstructed $\decay{B}{\Kp \ep \en (X)}$ decays, contaminate the region below the \Bp mass and are reduced using pion PID requirements. Decays mediated by $\decay{b}{c}$ transitions represent a significant background due to their large branching fractions and similar final states to the signal. 
The largest of these contributions arises from $\decay{B^+}{\Dzb(K^+ e^- \neueb) e^+ \neue}$ decays, where the kaon is misidentified as a pion. These are rejected with the requirement $m(\pi^+_{\to K} e^-)>m_{\Dz}$, where $m(\pi^+_{\to K} e^-)$ represents the invariant mass of the pion candidate and oppositely charged electron calculated when the pion candidate is assigned the kaon mass, and $m_{\Dz}$ is the known $\Dz$ mass~\cite{PDG2024}. 
The contribution from the $\decay{B^+}{\Dzb(\pi^+ e^- \neueb) e^+ \neue}$ decay is also suppressed by this veto, and further reduced by the requirements introduced to suppress combinatorial background. Similar decays where a hadron is misidentified as an electron, such as $\decay{B^+}{\Dzb (h^+ \pi^-) e^+ \neue}$, are suppressed by vetoing candidates with $m(\pi^+ e^-_{\to h})$ around the  known $\Dz$ mass.

Residual contamination from any of these backgrounds is estimated using simulation. The expected contribution from misidentified $\decay{B^+}{h^+h^-h^+}$ and $\decay{B^+}{h^+ h^- \ep \neue}$ backgrounds is also validated using a data-driven method developed for a similar LHCb analysis of $\decay{\bquark}{\squark \ep\en}$ decays~\cite{LHCb-PAPER-2022-045}, based on a control sample selected by inverting the PID requirements.

Combinatorial background is suppressed using a boosted decision tree (BDT) classifier~\cite{Breiman,AdaBoost}, implemented in the TMVA toolkit~\cite{TMVA4}. The BDT is trained with simulated \BuTopiee decays as a proxy for signal, while background is represented by data in the lower and upper mass sidebands, defined by $\mpiee < 5.0\gevcc$ and $\mpiee > 5.5\gevcc$, respectively. The training features include kinematic and topological properties of the final-state particles, the dielectron candidate and the reconstructed $B^+$ meson, as well as the fit quality of the $B^+$ decay vertex and its displacement with respect to its associated PV, which is defined as the PV with which the candidate has the smallest impact parameter significance. Two different BDT-output regions, referred to as LowBDT and HighBDT, are retained for the final analysis to help control the modelling of the combinatorial background in the final fit to data. The definition of the BDT regions is optimised using simulated pseudoexperiments in each $\qsq$ and bremsstrahlung category to maximise the signal significance, giving eight distinct signal regions.

The branching fraction of the \BuTopiee decay, extracted using a simultaneous maximum-likelihood fit to the eight signal regions, is related to the signal yield and signal efficiency in each region by 
\begin{align}
\BR(\BuTopiee) =& \frac{N^i(\BuTopiee)}{N(\BuToKJpsiee)} \cdot \frac{\epsilon(\BuToKJpsiee)}{\epsilon^i(\BuTopiee)} \nonumber \\
& \cdot \BR(\BuToKJpsi) \cdot \BR(\jpsi \to e^+ e^-),
\end{align}
where $N$ and $\epsilon$ are the observed yield and reconstruction and selection efficiency for the corresponding decay mode, respectively. The index $i$ refers to each of the eight signal categories defined according to the \qsq region, the bremsstrahlung recovery status and the value of the BDT output. The branching fractions of \BuToKJpsi and $\jpsi \to e^+ e^-$ decays are taken from Ref.~\cite{PDG2024}.

The efficiencies to reconstruct and select the signal and normalisation modes are estimated using simulated samples that are calibrated using control samples to better represent the data.  Trigger efficiencies are calibrated using a data-driven method applied to the normalisation mode~\cite{LHCb-PUB-2014-039}. The efficiency of the PID requirements is obtained using samples of decays that can be efficiently selected in data without the use of PID criteria~\cite{LHCb-PUB-2016-021,LHCb-DP-2018-001}, and applied as per-event weights to the simulated samples. Finally, the mass and \qsq resolutions are smeared to match those observed in data.

\begin{figure}
\centering
\includegraphics[width=0.5\textwidth]{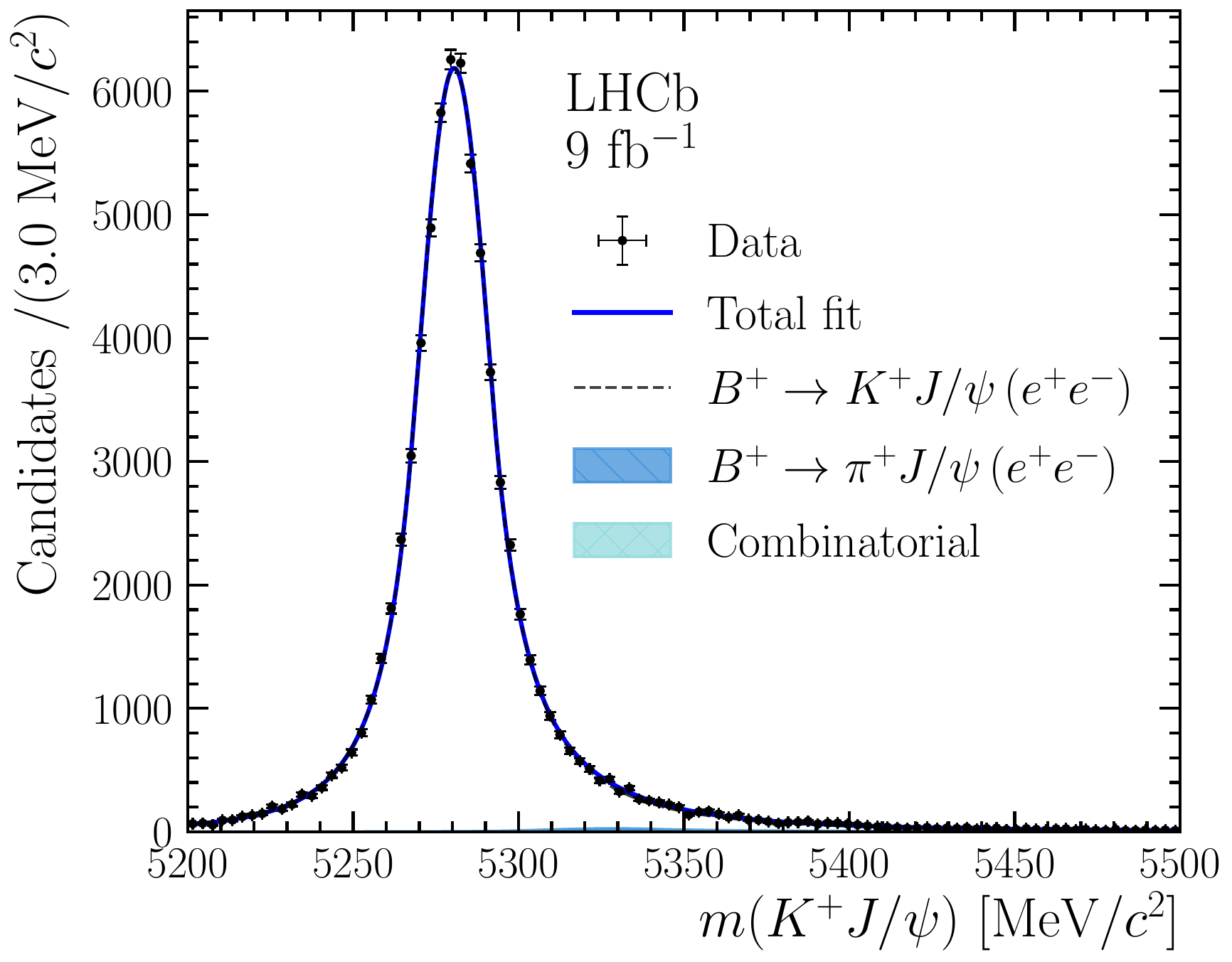}
\caption{Invariant-mass distribution of  selected \BuToKJpsiee candidates. The results of the fit are also shown. Some fit components are too small to be visible.\label{fig:KJpsiee_fit}}
\end{figure}

The yield of the normalisation mode is determined from a maximum-likelihood fit to the invariant-mass distribution of the selected \BuToKJpsiee candidates, which is computed after constraining the dielectron mass to that of the \jpsi meson~\cite{PDG2024} to improve the resolution. The \BuToKJpsiee signal is described by a double-sided Crystal Ball function~\cite{Skwarnicki:1986xj} with tail parameters fixed from simulation. The peak position and width parameter are allowed to vary in the fit. Background from the Cabibbo-suppressed \BuTopiJpsiee decay is modelled using simulation and its normalisation relative to the \BuToKJpsiee component is constrained to the expected value. An exponential function with free yield and slope parameters is used to describe the combinatorial background. The invariant-mass distribution of the \BuToKJpsiee sample, together with the results from the fit, is shown in \figref{fig:KJpsiee_fit}.

The efficiency estimation procedure and the normalisation mode yield extraction are validated using fits to \BuTopiJpsiee decays, where the final state hadron must satisfy the pion PID requirements. The ratio of branching fractions $\BR(\BuTopiJpsi)/\BR(\BuToKJpsi)$ is evaluated with and without a \jpsi mass constraint applied to the dielectron pair in \BuTopiJpsiee decays, with both results in agreement with the world average \cite{PDG2024} within one standard deviation.

The branching fraction of the \BuTopiee decay is extracted from an extended simultaneous unbinned maximum-likelihood fit to the invariant-mass spectra of the selected signal candidates in the eight different categories introduced above. The invariant-mass distribution of the signal is described by a double-sided Crystal Ball function with tail parameters determined from simulation. The peak position and width are fixed to the values obtained from fits to the normalisation mode data without the dielectron mass constraint. 

Contributions from residual physical backgrounds are parameterised using dedicated simulated samples and their yields are constrained to the corresponding expectation values. These include partially reconstructed $b$-hadron decays via \decay{b}{s e^+ e^-} and \decay{b}{d e^+ e^-} transitions, such as \BdToKstpzee or \BuTorhopzee, where one of the hadrons is not reconstructed. Some of these decays have never been observed, hence their yields are estimated from the branching fractions of the corresponding \decay{b}{s \mu^+ \mu^-} or \decay{b}{d \mu^+ \mu^-} transitions, assuming lepton universality, and a systematic uncertainty is evaluated to account for this assumption. A small fraction of decays involving charmonium resonances, \ie \BuToXJpsiee and \BuToXpsitwosee, can evade the \qsq and \qsqtrack vetoes, and their contributions are included with their yield fixed to the expected value. Finally, two sources of background from misidentified particles are considered.
The \BuToKee decay, where the kaon is misidentified as a pion, populates the region immediately below the signal peak. In contrast, fully hadronic \decay{B^+}{h^+ h^- h^+} decays, where two hadrons are misidentified as electrons, tend to accumulate under the signal region, but only in the NoBrem category, where the absence of bremsstrahlung recovery reduces the power of the PID requirements. The yields of these contributions are fixed to the expected values based on their branching fractions and the efficiency of the PID selection.

\begin{figure}
\centering
\includegraphics[width=\textwidth]{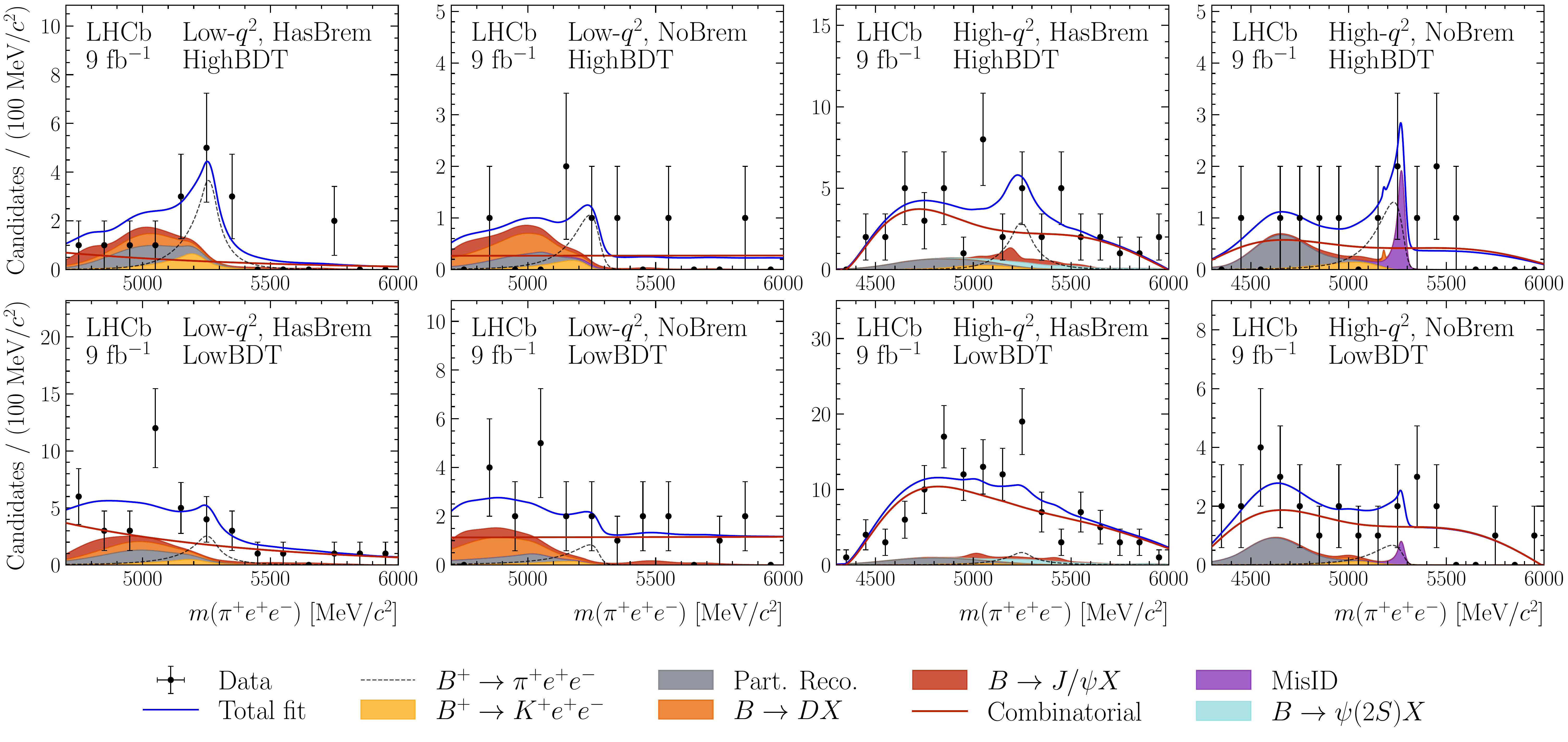}
\caption{Invariant-mass distribution of  selected \BuTopiee candidates in each of the eight signal categories. The results of the fit are also shown. \label{fig:piee_fit}}
\end{figure}

The combinatorial background is modelled by an exponential function in the low-\qsq regions. In the high-\qsq regions, the restricted phase space induces a suppression of combinatorial background at low invariant mass, which modifies the shape of this component from an exponential distribution. This effect is described using the  phase-space model introduced in Ref.~\cite{LHCb-PAPER-2024-056}. The effect of the rest of the selection is incorporated by multiplying the phase-space model by sculpting functions determined from two control samples in data. The first one is a mixed sample in which pion and electron candidates from different $pp$ collisions are combined to form a fake $B^+$ candidate. The mixed sample is used to validate the description of the phase-space model and to determine the sculpting associated to the kinematic vetoes. The second sample comprises \decay{B^+}{\pi^+ \mu^{\pm} e^{\mp}} candidates in data, and is used to determine the sculpting induced by the BDT selection. The combinatorial yields are allowed to vary independently in each of the eight categories, while the shape parameters are shared between the corresponding LowBDT and HighBDT regions. The invariant-mass distributions of the selected \BuTopiee candidates in each of the eight categories are shown in \figref{fig:piee_fit}, together with the results of the fit.

Systematic uncertainties arise from the evaluation of the efficiencies of signal and normalisation modes, from the modelling of the invariant-mass distributions and from the effect of possible residual backgrounds not included in the baseline fit model. Systematic uncertainties are evaluated by generating pseudoexperiments with alternative configurations of the fit model or alternative values of the constrained parameters. The effect of residual differences between data and simulation on the calculated efficiencies is estimated by recomputing the efficiencies without each of the corrections applied to simulation. The model used to describe the combinatorial background in the high-\qsq region~\cite{LHCb-PAPER-2024-056} uses an exponential distribution to which sculpting functions derived from control samples are applied. An alternative model, extracted from a fit to candidates with an inverted BDT selection, is used to evaluate the choice of an exponential distribution. In addition, the statistical variation on the sculpting effects is estimated by bootstrapping the corresponding control samples and recomputing the sculpting functions. A large number of potential physical backgrounds are investigated using simulated samples. In some cases, the limited size of these simulation samples induces a large uncertainty on the expected background contribution, and the parametrisation of these components in the fit is impractical. These include the semileptonic decays $\decay{B^+}{\rho(770)^0( \pi^+ \pi^-) e^+ \neue}$ and $\decay{B^+}{\Dzb(\pi^+ e^- \neueb) e^+ \neue}$, as well as $b$-hadron decays with photons in the final state that convert into electron-positron pairs in the detector, one of which is selected as one of the electron candidates. The systematic uncertainty associated with ignoring these background components is estimated by generating pseudoexperiments with these additional backgrounds and evaluating the induced bias in the branching fraction when the baseline model is used in the fit. For those backgrounds without a measured branching fraction, the estimated yields are varied by 100\% and the associated bias in the result is taken as the systematic uncertainty. A summary of the systematic uncertainties is shown in \tabref{tab:systematics}. The total uncertainty is obtained by adding all contributions in quadrature.

\begin{table}
\centering
\renewcommand{\arraystretch}{1.2}
\caption{\small Relative systematic uncertainties on the \BuTopiee branching fraction measurement in the low- and high-\qsq regions, and on the integrated measurement.\label{tab:systematics}}
\begin{tabular}{l|c|c|c}
\hline
Source & Low-\qsq [\%] & High-\qsq [\%] & Integrated [\%] \\
\hline
Efficiency estimation & $13.4$ & $12.4$ & $10.7$ \\
Combinatorial model & $\phantom{0}1.2$ & $54.2$ & $\phantom{0}7.4$    \\
Physical backgrounds & $\phantom{0}7.1$ & $26.3$ & $\phantom{0}3.0$ \\
\hline
Total uncorrelated & $15.2$ & $61.5$ & $13.3$ \\
\hline
$\BR(\BuToKJpsiee)$ & $\phantom{0}1.9$ & $\phantom{0}1.9$ & $\phantom{0}1.9$ \\
\hline
Total & $15.3$ & $61.5$ & $13.4$ \\
\hline
\end{tabular}
\end{table}

The branching fraction is determined to be
\begin{equation*}
\BR(\BuTopiee) = (2.4\,{}^{+0.9}_{-0.8} \,{}^{+0.4}_{-0.2}) \times 10^{-8},
\end{equation*}
where the first uncertainties are statistical and the second are systematic.  This result should be interpreted as a measurement of the total branching fraction in the entire \qsq range accessible, including the regions of \qsq not measured in this analysis. The extrapolation to the full \qsq range is performed using \BuTopiee simulation in which $c\bar{c}$ resonances are not included~\cite{Ball:2004ye}.
The significance of the signal contribution is estimated using Wilks' theorem~\cite{Wilks:1938dza} and found to be $3.2\sigma$, taking into account systematic uncertainties.
This constitutes the first evidence for the \BuTopiee decay. The profile-likelihood function is shown in the End Matter.

The fit is also repeated by allowing the branching fraction to vary independently in the low-\qsq and high-\qsq regions, which are determined to be
\begin{equation*}
\begin{aligned}
\BR(\BuTopiee)^{[0.045, 6] \gevgevcccc} &= (0.76\,{}^{+0.30}_{-0.27} \,{}^{+0.14}_{-0.08}) \times 10^{-8},\\
\BR(\BuTopiee)^{[15, 25] \gevgevcccc} &= (0.40\,{}^{+0.52}_{-0.45} \,{}^{+0.24}_{-0.22}) \times 10^{-8}.
\end{aligned}
\end{equation*}
Since 24.6\% and 34.5\% of simulated signal candidates are produced in the low- and high-\qsq regions respectively, these results are consistent with each other within one standard deviation. 

The significance of the signal in these two regions is $3.2\sigma$ and $0.8\sigma$, respectively. As no signal is observed in the high-\qsq region, an upper limit to the branching fraction is computed using the CL$_{\rm s}$ method~\cite{CLs} to be $\BR(\BuTopiee)^{[15, 25]\gevgevcccc}<1.25\, (1.49) \times 10^{-8}$ at 90\% (95\%) confidence level. 
All of the results above are in good agreement with the SM expectation~\cite{piee_SM}. 

In summary, a search for the \BuTopiee decay has been performed using data collected by the LHCb experiment during Run~1 and Run~2 of the LHC, corresponding to an integrated luminosity of 9\invfb. Evidence for the signal is found with a significance of $3.2\sigma$ and the branching fraction is determined for the first time. The result is in good agreement with the SM expectation and represents the first evidence for a \decay{b}{d e^+ e^-} transition. 


\section*{Acknowledgements}
%
%
\noindent We express our gratitude to our colleagues in the CERN
accelerator departments for the excellent performance of the LHC. We
thank the technical and administrative staff at the LHCb
institutes.
We acknowledge support from CERN and from the national agencies:
ARC (Australia);
CAPES, CNPq, FAPERJ and FINEP (Brazil); 
MOST and NSFC (China); 
CNRS/IN2P3 and CEA (France);  
BMFTR, DFG and MPG (Germany);
INFN (Italy); 
NWO (Netherlands); 
MNiSW and NCN (Poland); 
MEC/IFA (Romania); 
MICIU and AEI (Spain);
SNSF and SER (Switzerland); 
NASU (Ukraine); 
STFC (United Kingdom); 
DOE NP and NSF (USA).
We acknowledge the computing resources that are provided by ARDC (Australia), 
CBPF (Brazil),
CERN, 
IHEP and LZU (China),
IN2P3 (France), 
KIT and DESY (Germany), 
INFN (Italy), 
SURF (Netherlands),
Polish WLCG (Poland),
IFIN-HH (Romania), 
PIC (Spain), CSCS (Switzerland), 
GridPP (United Kingdom),
and NSF (USA).  
We are indebted to the communities behind the multiple open-source
software packages on which we depend.
Individual groups or members have received support from
RTP (Australia), 
Key Research Program of Frontier Sciences of CAS, CAS PIFI, CAS CCEPP (China); 
Minciencias (Colombia);
EPLANET, Marie Sk\l{}odowska-Curie Actions, ERC and NextGenerationEU (European Union);
A*MIDEX, ANR, IPhU and Labex P2IO, and R\'{e}gion Auvergne-Rh\^{o}ne-Alpes (France);
Alexander-von-Humboldt Foundation (Germany);
ICSC (Italy); 
Severo Ochoa and Mar\'ia de Maeztu Units of Excellence, GVA, XuntaGal, GENCAT, InTalent-Inditex and Prog.~Atracci\'on Talento CM (Spain);
the Leverhulme Trust, the Royal Society and UKRI (United Kingdom).

\clearpage

\section*{End matter}

\begin{figure}[h]
\centering
\includegraphics[width=0.5\textwidth]{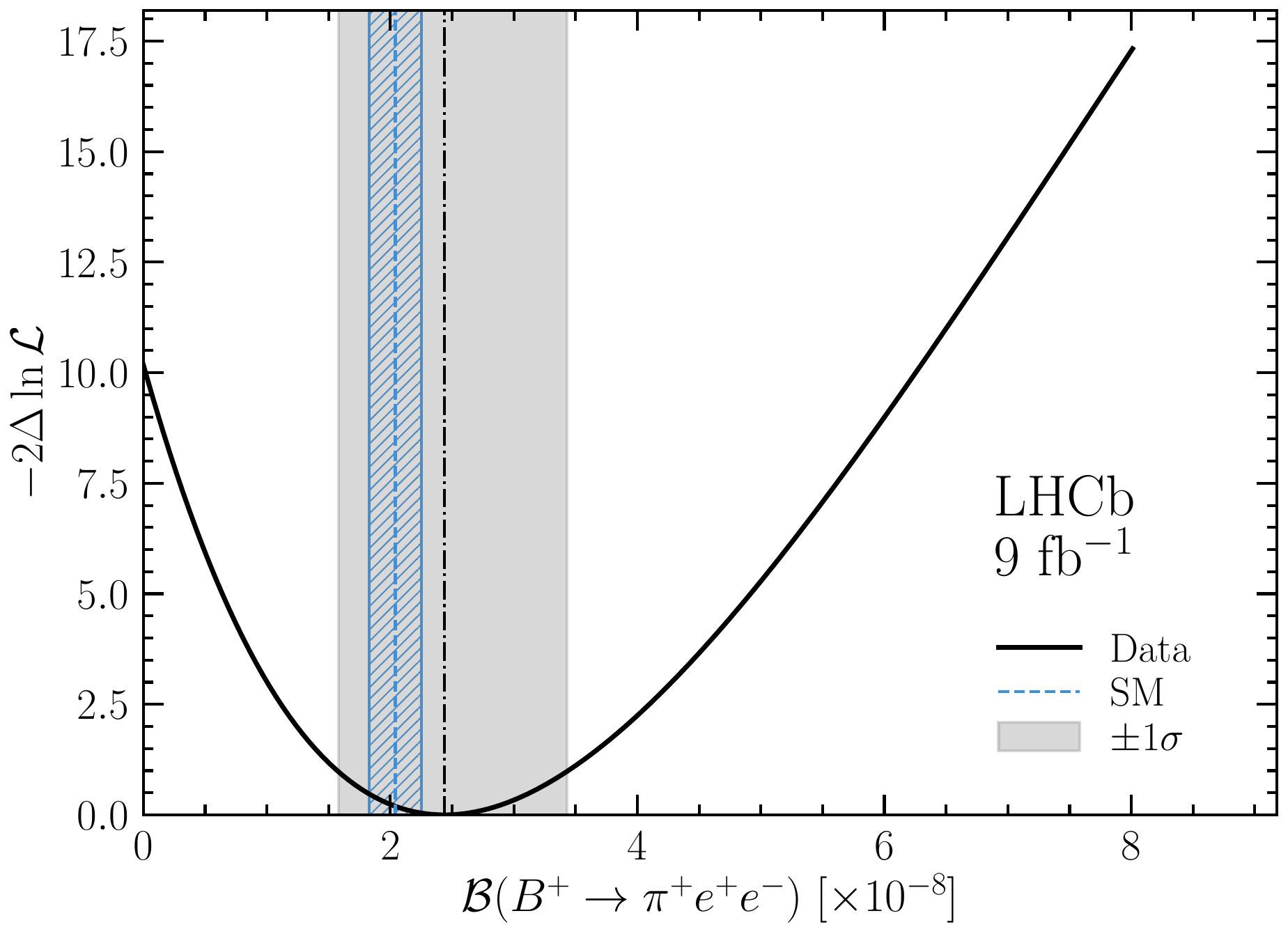}
\caption{Profile likelihood as a function of the \BuTopiee branching fraction for the integrated \qsq region. The blue band indicates the SM prediction~\cite{piee_SM}.\label{fig:likelihood_integrated}}
\end{figure}

\begin{figure}[h]
\centering
\includegraphics[width=0.5\textwidth]{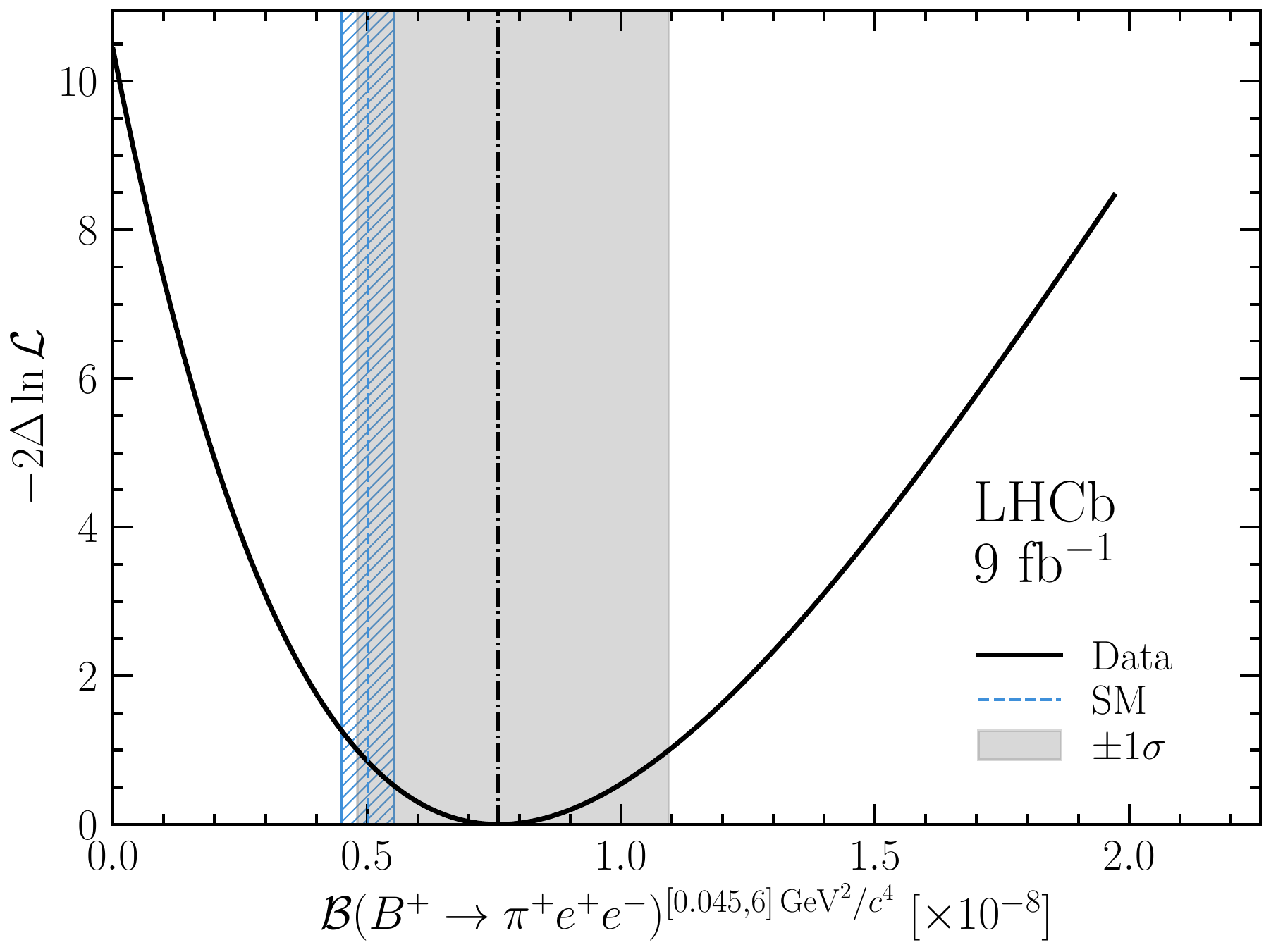}\hfill
\includegraphics[width=0.5\textwidth]{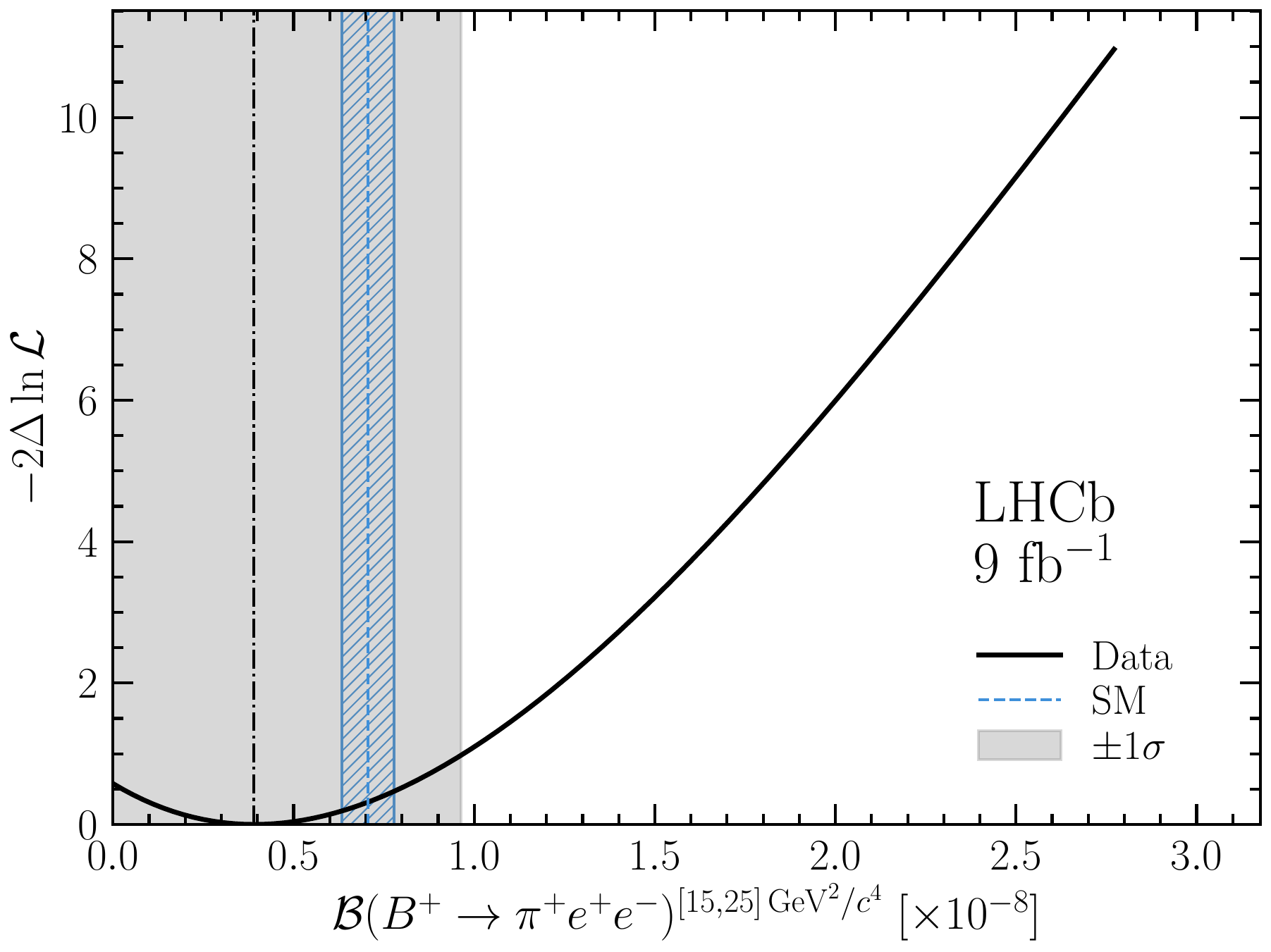}
\caption{Profile likelihood as a function of the \BuTopiee branching fraction for the  (left) low-\qsq  and (right) high-\qsq regions. The blue band indicates the SM prediction~\cite{piee_SM}.}\label{fig:likelihood_qsq}
\end{figure}

The profile likelihoods provide the complete statistical information for the fitted branching fractions, including any non-Gaussian behavior, and therefore facilitate reinterpretation and combination with future measurements. The profile likelihood as a function of the \BuTopiee branching fraction is shown in \figref{fig:likelihood_integrated} for the integrated \qsq region and in \figref{fig:likelihood_qsq} for the two separate \qsq regions.

\clearpage

\addcontentsline{toc}{section}{References}
\bibliographystyle{LHCb}
\bibliography{main,standard,LHCb-PAPER,LHCb-CONF,LHCb-DP,LHCb-TDR}

\newpage
\centerline
{\large\bf LHCb collaboration}
\begin
{flushleft}
\small
R.~Aaij$^{38}$\lhcborcid{0000-0003-0533-1952},
A.S.W.~Abdelmotteleb$^{58}$\lhcborcid{0000-0001-7905-0542},
C.~Abellan~Beteta$^{52}$\lhcborcid{0009-0009-0869-6798},
F.~Abudin\'en$^{58}$\lhcborcid{0000-0002-6737-3528},
T.~Ackernley$^{62}$\lhcborcid{0000-0002-5951-3498},
A. A. ~Adefisoye$^{70}$\lhcborcid{0000-0003-2448-1550},
B.~Adeva$^{48}$\lhcborcid{0000-0001-9756-3712},
M.~Adinolfi$^{56}$\lhcborcid{0000-0002-1326-1264},
P.~Adlarson$^{86}$\lhcborcid{0000-0001-6280-3851},
C.~Agapopoulou$^{14}$\lhcborcid{0000-0002-2368-0147}
% ...

{\footnotesize \it

% ...
$^{14}$Universit{\'e} Paris-Saclay, CNRS/IN2P3, IJCLab, Orsay, France\\
% ...
$^{38}$Nikhef National Institute for Subatomic Physics, Amsterdam, Netherlands\\
% ...
$^{48}$Instituto Galego de F{\'\i}sica de Altas Enerx{\'\i}as (IGFAE), Universidade de Santiago de Compostela, Santiago de Compostela, Spain\\
% ...
$^{52}$Physik-Institut, Universit{\"a}t Z{\"u}rich, Z{\"u}rich, Switzerland\\
% ...
$^{56}$H.H. Wills Physics Laboratory, University of Bristol, Bristol, United Kingdom\\
% ...
$^{58}$Department of Physics, University of Warwick, Coventry, United Kingdom\\
% ...
$^{61}$School of Physics and Astronomy, University of Glasgow, Glasgow, United Kingdom\\
$^{62}$Oliver Lodge Laboratory, University of Liverpool, Liverpool, United Kingdom\\
% ...
$^{70}$Syracuse University, Syracuse, NY, United States\\
% ...
$^{86}$Department of Physics and Astronomy, Uppsala University, Uppsala, Sweden, associated to $^{61}$\\
% ...
}
\end{flushleft}

\end{document}